\documentclass[a4paper,fleqn,usenatbib,useAMS]{mnras}

\usepackage{graphicx}	
\usepackage{amsmath}	
\usepackage{amssymb}	
\usepackage{multicol}        
\usepackage{bm}		
\usepackage{pdflscape}	

\usepackage[T1]{fontenc}
\usepackage{ae,aecompl}
\usepackage{mathptmx}

\usepackage{hyperref}
\PassOptionsToPackage{hyphens}{url}\usepackage{hyperref}
\makeatletter
\g@addto@macro{\UrlBreaks}{\UrlOrds}
\makeatother

\newcommand{\kms}{\,km\,s$^{-1}$} 

\title{The separate effect of halo mass and stellar mass on the evolution of massive disk galaxies}

\author[S. Zhou]{Shuang Zhou$^{1,2}$\thanks{Contact e-mail: \href{mailto:Shuang.Zhou@inaf.it}{Shuang.Zhou@inaf.it}},
Alfonso Arag{\'o}n-Salamanca$^{1}$\thanks{Contact e-mail: \href{mailto:Alfonso.Aragon@nottingham.ac.uk}{Alfonso.Aragon@nottingham.ac.uk}},
Michael Merrifield$^{1}$
\\
$^{1}$School of Physics \& Astronomy, University of Nottingham, University Park, Nottingham, NG7 2RD, UK\\
$^{2}$INAF-Osservatorio Astronomico di Brera, via Brera 28, I-20121
Milano, Italy\\
}
\date{Last updated ???; in original form ???}

\pubyear{2022}

\begin{document}

\label{firstpage}
\pagerange{\pageref{firstpage}--\pageref{lastpage}}
\maketitle

\begin{abstract}
We analyse a sample of massive disk galaxies selected from the SDSS-IV/MaNGA survey to investigate how the evolution of these galaxies depends on their stellar and halo masses.
We applied a semi-analytic spectral fitting approach to the data from different regions in the galaxies to derive several of their key physical properties. From the best-fit model results, together with direct observables such as morphology, colour, and the Mgb/$\langle$Fe$\rangle$ index ratio measured within $1 R_{\rm e}$, we find that for central galaxies both their stellar and halo masses have a significant influence in their evolution. For a given halo mass, galaxies with higher stellar mass accumulate their stellar mass and become chemically enriched earlier than those with smaller stellar mass.  Furthermore, at a given stellar mass, galaxies living in more massive halos have longer star-formation timescales and are delayed in becoming chemically enriched. In contrast, the evolution of massive satellite galaxies is mostly determined by their stellar mass.  The results indicate that both the assembled halo mass and the halo assembly history impact the evolution of central galaxies. Our spatially resolved analysis indicates that only the galaxy properties in the central region ($0.0$--$0.5 R_{\rm e}$) show the dependencies described above. This fact supports a halo-driven formation scenario since the galaxies' central regions are more likely to contain old stars formed along with the halo itself,  keeping a memory of the halo formation process.

\end{abstract}



\section{Introduction}
In the current $\Lambda$CDM paradigm, galaxies form at the centre of dark matter halos \citep[e.g.][]{White1978}. It is thus perhaps unsurprising that a galaxy's evolution would correlate, to first-order, with the growth of the dark matter halo \citep[e.g.][]{White1978,Blumenthal1984}. The formation and evolution of dark matter halos are mainly driven by gravitational forces, which are relatively well understood through large N-body simulations \citep[e.g.][]{Springel2005}. Yet, galaxies are complex systems not regulated by dark matter alone; they are also affected by a variety of baryonic processes. The physics involved in these processes is highly non-linear and remains to be fully understood. It is therefore very important to investigate, simultaneously, how the physical properties of galaxies and their evolution are driven by their dark matter halos (i.e., the galaxy-halo connection), together with their dependence on physical properties connected with the baryonic physics. 

Searching for a link between the evolution of galaxies and their host dark matter halos has been challenging because the properties of the dark matter halos are often not directly observable. It has long been known that galaxies in high-density regions such as groups and clusters are more likely to be red and dead  \citep[e.g.][]{Oemler1974,Dressler1980,Postman1984}. More recent observations through  large surveys such as the Sloan Digital Sky Survey (SDSS,\citealt{York2000})  have revealed that satellite galaxies in groups and clusters --especially the low-mass ones--
can have their star-formation quenched by the dense environment \citep[e.g.][]{Pasquali2010,Peng2012,Wetzel2012,Wetzel2013}. For massive galaxies, however, being a central or satellite galaxy does not seem to alter their evolution paths. The consensus from a number of recent investigations is that the 
stellar populations of massive galaxies are largely determined by internal galaxy properties rather than the environment, which is related to the properties of their halos \citep[e.g.][]{Peng2012,Greene2015,Scott2017,Zhou2022environment}.

Results from large surveys such as SDSS also reveal that, of all the internal galaxy properties that may drive the evolution of massive galaxies, their stellar mass and/or central velocity dispersion are likely to be the most important  ones. More massive galaxies are generally found to be older and more metal-rich, with redder colours and earlier morphological types \citep[e.g.][]{Kauffmann2003,Gallazzi2005,Thomas2005}. In particular, it is found that the more massive galaxies, especially early types, are more enhanced in their [$\alpha$/Fe] ratios \citep[e.g.][]{Jorgensen1999,Thomas2005,Gallazzi2006,McDermid2015}. Such a ratio can serve as a proxy for the timescale of the galaxies' star formation: a galaxy with an early truncated star formation history will end up with enhanced [$\alpha$/Fe] ratio compared to galaxies with more extended star formation histories \citep{Worthey1994,Thomas2005}. In this sense, galaxies of higher present-day stellar masses have built up their stellar mass faster than lower-mass ones; this so-called ‘downsizing’ phenomenon provides important constraints to galaxy formation models.

However, it remains a puzzle whether ‘downsizing’ is solely driven by the stellar mass of the galaxies, or if it is a reflection of the different assembly histories of their dark matter halos. \cite{Scholz-Diaz2022} addressed the question by analyzing a sample of nearby central galaxies from SDSS and investigating the correlation between their stellar population properties and the properties of their host halos. They found that the ages and metallicities of galaxies are not fully determined by stellar mass -- at fixed stellar mass, the halo mass plays a secondary yet noticeable role so that galaxies in more massive halos tend to be younger and more metal-poor. Similar conclusions are reached by the work of \cite{Oyarzun2022}, in which IFU data from the Mapping Nearby Galaxies at Apache Point Observatory (MaNGA) survey are used to assert the influence of halo mass on stellar populations of passive central galaxies. From simulations, \cite{Wang2023}
find that central galaxies in more massive halos have lower gas-phase metallicity at high redshift than those in low-mass halos due to the long-lasting accretion of low-metallicity gas. Such an effect is used to explain the low metallicity of cluster galaxies found at high redshift \citep[e.g.][]{Wang2022}. These studies open a new window to investigating how the properties of dark matter halos can impact the evolution of galaxies. However, most of these studies are based on simple age or metallicity averages and are limited to single-fibre observation or specific types of galaxies. A full description of the spatially resolved past history of a large and well-defined sample of galaxies will provide key information on how galaxy halo and stellar masses shaped their evolution.

Fortunately, we are now in a position to investigate these physical processes in more detail. In recent work (\citealt{Zhou2022}, hereafter Paper~I), we have shown that it is possible to go beyond the simple averaged age/metallicity to reach a full description of the stellar mass accumulation and chemical enrichment histories of galaxies using a `semi-analytic' spectral fitting approach. This approach fits chemical evolution models directly to the spectra to determine physical parameters that characterise the timescales of star formation and the rates of gas accretion and loss, which can be more fundamentally related to the formation and evolution of the galaxies within their dark matter halos. From the observational perspective, the MaNGA survey \citep{Bundy2015} provides high-quality spatially resolved spectra for 
a well-defined sample of over 10,000 galaxies. The large sample size allows us to apply different grouping strategies to separate the effects from stellar mass and halo mass, while the spatially resolved power of MaNGA helps to pin down the effect on different regions, which may reflect various physical processes. Combining our advanced analysis tool with the large dataset, we expect to reveal in more detail how a galaxy's evolution, including the evolution of its star formation activity and chemical composition, can be shaped by its stellar mass and halo mass.

We thus set out the analysis and organize this paper as follows. We present the data and galaxy sample in \S\ref{sec:data}. A brief introduction to our `semi-analytic' spectral fitting process and how we implement it to the MaNGA dataset is presented in \S\ref{sec:analysis}. We then investigate the inferred evolutionary parameters and their correlation with galaxy and halo properties in \S\ref{sec:results}. Finally, we summarise our key results in \S\ref{sec:summary}.Throughout this work we use a standard $\Lambda$CDM cosmology with $\Omega_{\Lambda}=0.7$, 
$\Omega_{\rm M}=0.3$ and $H_0$=70\kms Mpc$^{-1}$.

\section{The data}
\label{sec:data}
In order to investigate the effect of both stellar mass and halo mass on the evolution of massive disk galaxies both globally and spatially resolved, we make use of the IFU data available from the SDSS-IV/MaNGA survey. In this section, we give a brief introduction to the survey and then describe our sample selection and data reduction processes.

\subsection{MaNGA}
MaNGA was part of the fourth generation of SDSS (SDSS-IV, \citealt{Blanton2017}). This survey provided  spatially-resolved high-quality spectroscopy data for over 10,000 nearby (redshift $0.01<z<0.15$) galaxies \citep{Yana2016}, covering the  stellar mass range $5\times10^8 h^{-2}{M}_{\odot} \leq M_*\leq 3 \times 10^{11} 
h^{-2}{M}_{\odot}$ \citep{Wake2017}.
The MaNGA IFU covers at least 1.5 effective radii for all the target galaxies \citep{Law2015}. The spectra were obtained with the two dual-channel BOSS spectrographs \citep{smee2013} mounted on the Sloan 2.5\,m telescope \citep{Gunn2006}, covering $3600-10300${\AA} in wavelength with 
intermediate spectral resolution ($R\sim2000$, \citealt{Drory2015}). MaNGA used a Data Reduction Pipeline (DRP; \citealt{Law2016}) to produce science-ready spectra with flux calibration better than $5\%$ across most of the wavelength range \citep{Yanb2016}. In addition, a Data Analysis Pipeline (DAP;\citealt{Westfall2019,Belfiore2019}) was used to process these reduced data, providing various useful data products including stellar kinematics, emission-line properties and spectral indices. Moreover, the MaNGA collaboration has carried out very extensive analysis and research with the data, and some of the outputs of that work have been released as value-added-catalogues (VACs); we will make use of a number of these results in this paper.

\subsection{Sample selection and data preparation}
We start from the parent sample of disk galaxies defined in Paper I. This parent sample consists of 2560 galaxies selected from the MaNGA survey. In a previous paper \citep{Zhou2022environment} we explored explicitly how the evolution of low-mass galaxies is affected by the environment. We will now focus now on more massive galaxies, with a stellar low mass limit of $10^{10.4}{\rm M_{\odot}}$. To obtain the group properties of these galaxies, we cross-match this sample with the group catalogue of \cite{Yang2007}, from which we identify massive central and satellite galaxies, as well as the group halo masses of these galaxies.

\begin{figure}
    \centering
    \includegraphics[width=0.4\textwidth]{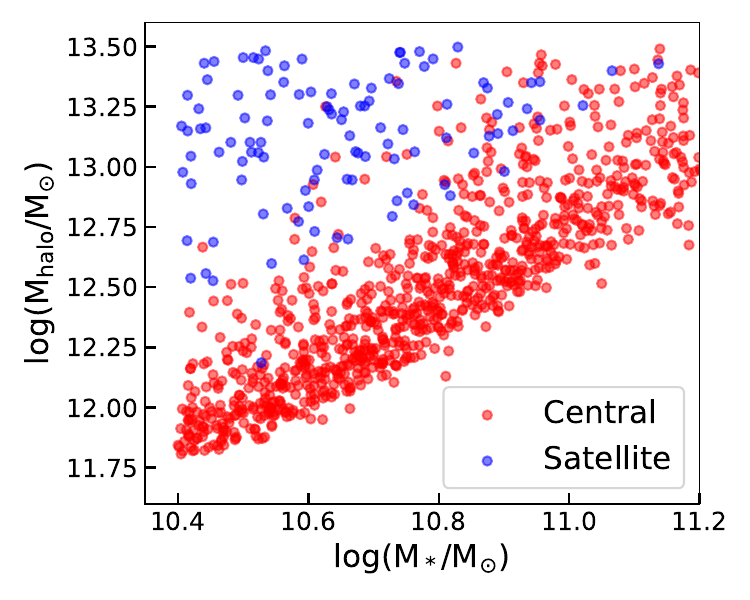}\\
     \caption{The group halo mass- stellar mass relation of our sample galaxies. Central galaxies are shown in red and satellites in blue. 
     }
     \label{fig:galaxy_sample}
\end{figure}

To avoid the most extreme environmental effects, we limit our investigation to galaxies in groups and not clusters. Hence, the host halo masses are selected to be $\lesssim 10^{13.5}{\rm M_{\odot}}$. After this selection, we end up with a sample of 972 massive central disk galaxies and 113 massive satellites. We show their group mass as a function of stellar mass in \autoref{fig:galaxy_sample}. As expected, the stellar masses of central galaxies correlate well with the group halo mass, while such a relation is absent for satellite galaxies. 

We note that there are several alternative group catalogues available for SDSS galaxies such as the one from \cite{Tinker2021}. Discussing the differences between alternative group catalogues goes beyond the scope of this work. However, as a consistency check, we have performed a similar analysis to the one described below using the group catalogue from \cite{Tinker2021} and find generally the same results as the ones using the \cite{Yang2007} catalogue. Similar analysis carried out by \cite{Oyarzun2022} also indicates that the results of their analysis were robust against group catalogue choice. We thus decided to use the catalogue from \cite{Yang2007} throughout this paper to facilitate comparisons with previous published work. In what follows we will focus on the sample constructed above to discuss the effect of stellar mass and halo mass on the evolution of galaxies.

As mentioned above, observational data for our sample galaxies comes from the MaNGA survey. The raw data cubes provided by the MaNGA DRP contain spectra for individual spaxels with typical signal-to-noise ratios (SNR) in the $r$-band  $\sim4$--$8${\,\AA}$^{-1}$ at the outskirts of galaxies \citep{Law2016}, which is not sufficient for robust stellar-population (specially metallicity) analysis. We thus stack spaxels to obtain spectra with the required SNR. In this work, we intend to investigate both the global and spatially-resolved properties of the galaxies. To this end, we perform two types of stacking. For the global properties, we use all the spaxels  within $1R_{\rm e}$, with $R_{\rm e}$ being the effective radius of each galaxy. This achieves sufficient SNR to calculate the galaxies' integrated properties with a high degree of confidence. To investigate the stellar population properties in a spatially-resolve way, we need to reach a compromise between the required SNR and spatial resolution. This is achieved by dividing each galaxy into two regions: an inner region within $0.5R_{\rm e}$, and and outer region from $0.5R_{\rm e}$ to $1.5R_{\rm e}$, stacking the spaxels within each radial range. The stacking procedure is similar to the one used in\cite{Zhou2019}. The spectra of individual spaxels are shifted to the rest frame using the stellar kinematics information provided by the MaNGA DAP. The rest-frame spectra are then co-added together to obtain a single spectrum with high SNR. Note that adjacent spaxels from MaNGA datacubes are not fully independent; we therefore correct for co-variance between spaxels using the correction term from \cite{Westfall2019}. The stacking procedure yields spectra with a typical SNR (averaged over all wavelengths) of $\sim70$ per {\AA} for the $1R_{\rm e}$ stacks,  and $\sim50$ per {\AA} for the inner and outer regions. Such SNRs are suitable to derive reliable star-formation histories (SFHs) and chemical evolution histories (ChEHs) for the galaxies \citep{Zhou2022}.

Having obtained integrated spectra for the regions defined above, we derive some galaxy properties directly from the emission lines. These properties will provide additional constraints to our model. Firstly, we derive the current star-formation rate (SFR) of the relevant regions. This is achieved by obtaining the H${\alpha}$ flux measurements of every individual spaxel from the MaNGA DAP. DAP models simultaneously the continuum and emission lines for each MaNGA spectrum, providing continuum-subtracted emission line fluxes of various emission lines. For the H${\alpha}$ flux,  we correct for dust attenuation using the Balmer decrement assuming an intrinsic H${\alpha}$/H${\beta}$ ratio of 2.87 \citep{Osterbrock2006}. Dust-corrected H${\alpha}$ fluxes from all spaxels in each region are coadded to derive the total H${\alpha}$ flux and luminosity. This is then used to calculate the SFR following the calibration of \cite{Murphy2011} assuming a \cite{Chabrier2003} initial mass function (IMF):
\begin{equation}
\label{eq:sfr}
SFR({\rm M_{\odot}yr^{-1}})=5.37\times10^{-42}L_{\rm H\alpha}(\rm {erg\ s}^{-1}).
\end{equation} 

In addition to the SFR, we estimate the gas-phase metallicity of each region to provide further constraints to our models. To this end, we obtain measurements of four emission lines, i.e., \hbox{[O\,{\sc iii}]}$\lambda$5007, \hbox{[N\,{\sc ii}]}$\lambda$6584, H${\alpha}$, and  H${\beta}$ for every spaxel of the galaxy from the MaNGA DAP.  We then use the $O3N2$ indicator with the calibration of \cite{PP04} to estimate the gas-phase metallicity of each spaxel. The value for each region is obtained from the median of the corresponding spaxels. To 
facilitate the comparison between metallicities calculated from the chemical evolution model and the gas emission lines, we normalized all the gas-phase metallicities to the solar oxygen abundance. We use a solar
metallicity of $0.02$ and a solar oxygen abundance of $12+\log$(O/H)$\,= 8.83$ \citep{Anders1989}. These values are adopted for consistency with the settings of the \cite{BC03} SSP model used throughout this work (see \autoref{sec:fitting}). Obviously, caution is urged when comparing with results obtained with alternative calibrations.

\section{Analysis}
\label{sec:analysis}
The observational data, including the stacked spectra, SFR, and gas-phase metallicities are modelled using a semi-analytic spectral fit-
ting approach. This approach introduces a simple but flexible chemical evolution model 
to characterise the stellar mass accumulation and chemical evolution history of galaxies, which are combined with stellar population synthesis to produce model spectra that can be directly compared with observations. The model catches the main physical processes involved in galaxy evolution, and it is designed to be simple but flexible enough to produce model spectra that fit most kinds of galaxies. More details on the model and the tests performed to validate it can be found in Paper I. Here 
we will just briefly introduce the main ingredients of the approach.

\subsection{The chemical evolution model}
In the general picture, the main physics governing the star formation and chemical evolution of a galaxy includes gas inflow and outflow, as well as the star formation and chemical enrichment process that happens within the resulting gas reservoir. Taking into account all these processes, the gas mass evolution of a galaxy can be modelled  with the following equation:
\begin{equation}
\dot{M}_{\rm g}(t)=\dot{M}_{\rm in}(t)-\psi(t)+\dot{M}_{\rm re}(t)-\dot{M}_{\rm out}(t).
\end{equation}
The first term on the right-hand side represents how the gas falls into the galaxy. Such a process is often modelled with an exponentially decaying function 
\begin{equation}
\dot{M}_{\rm in}(t)=A e^{-(t-t_0)/\tau}, \ \ \ t > t_0.
\end{equation}
The second and third terms characterise the gas lost due to star formation and returned from dying stars respectively, while the final term represents the gas outflow process.  The star formation activity can be modelled with a linear Schmidt law \citep{Schmidt1959} that reads 
 \begin{equation}
\psi(t)=S\times M_{\rm g}(t),
\end{equation}
with $S$ being the star-formation efficiency.
We estimate this efficiency using the extended Schmidt law
proposed by \cite{Shi2011}, 
\begin{equation}
\label{eq:sfe}
    S (yr^{-1})=10^{-10.28\pm0.08} \left( \frac{\Sigma_*}{{\rm M_{\odot} pc^{-2}}} \right)^{0.48},
\end{equation}
where $\Sigma_{*}$ is the stellar mass surface density. For the investigation of global properties of the galaxy, we obtain an approximate $\Sigma_{*}$ using its current stellar mass and effective radius from the NSA so that $\Sigma_*=0.5\times M_*/(\pi R_{\rm e}^2)$. For the inner (0--0.5$\,R_{\rm e}$) and outer (0.5--1.5$\,R_{\rm e}$) regions, we make use of the stellar mass map of each galaxy obtained from the pip3D value-added catalogue to calculate the stellar mass in each region and derive their average  $\Sigma_{*}$.

The return of gas from dying stars is modelled with a simple constant mass return fraction of $R=0.3$, which means 30\% of the stellar mass formed in each generation will be returned to the ISM. This return is assumed to happen instantaneously when the stellar population is formed. The outflow term is traditionally modelled to have the outflow strength proportional to the star formation activity:
\begin{equation}
\label{eq:outflow}
\dot{M}_{\rm out}(t)=\lambda\psi(t),
\end{equation}
with a dimensionless quantity $\lambda$ being the so-called  `wind parameter'. In this work, following paper I, we allow the wind parameter to vary with time by simply turning it off at a time $t_{\rm cut}$, which is a free parameter during the fitting.
This cutoff represents when a galaxy's potential well becames too deep for the winds to escape and it allows a secondary star-formation episode during the evolution of the galaxy which provides additional flexibility in matching observations. 
Under all these assumptions, we derive the final equation that characterises the gas mass evolution as
\begin{equation}
\label{eq:massevo}
\dot{M}_{\rm g}(t)=
 \left\{\begin{array}{lr}
   A e^{-(t-t_0)/\tau}-S(1-R+\lambda) M_{\rm g}(t) &\text{(for $t<t_{\rm cut}$)}\\
   A e^{-(t-t_0)/\tau}-S(1-R) M_{\rm g}(t) &\text{(for $t>t_{\rm cut}$)}
   \end{array}.
   \right.
\end{equation}

We now turn to the metallicity evolution of this gas component. Throughout this work, we adopt an instantaneous mixing approximation, which assumes that the gas in a galaxy is always well mixed during its evolution.

In this case, the equations governing the chemical evolution can be written down as 
\begin{equation}
\label{eq:cheevo1_MANGA}
\begin{aligned}
\dot{M}_{Z}(t)=& Z_{\rm in}\dot{M}_{\rm in}(t)-Z_{\rm g}(t)(1-R)SM_{\rm g}(t) + y_Z(1-R)SM_{\rm g}(t)\\
& -Z_{\rm g}(t)\lambda SM_{\rm g}(t),
\end{aligned}
\end{equation}
 and 
\begin{equation}
\label{eq:cheevo2_MANGA}
\begin{aligned}
\dot{M}_{Z}(t)= Z_{\rm in}\dot{M}_{\rm in}(t)-Z_{\rm g}(t)(1-R)SM_{\rm g}(t) + y_Z(1-R)SM_{\rm g}(t),
\end{aligned}
\end{equation}
for $t<t_{\rm cut}$ and  $t>t_{\rm cut}$, respectively. In these equations, $Z_{\rm g}(t)$ is the metallicity of the gas so that $M_{Z}(t)\equiv M_{\rm g}\times Z_{\rm g}$ is the total mass of metals in the gas phase. The effect of the inflow is described by the first term, in which gas with an initial metallicity $Z_{\rm in}$ is falling into the system. It is often assumed that the infall gas is pristine so we always set $Z_{\rm in}=0$. 
The second term represents the gas mass locked up in long-lived stars, while the third term describes how the chemically-enriched gas from dying stars pollutes the ISM. A free parameter $y_Z$ is adopted to characterise the fraction of metal mass generated per stellar mass. Finally, the last term in \autoref{eq:cheevo1_MANGA} is the outflow term, representing the removal of metal-enriched gas.

\begin{table}
	\centering
	\caption{Priors of model parameters}
	\label{tab:paras}
	\begin{tabular}{lccr}
		\hline
		Parameter & Description & Prior range\\
		\hline
		$y_Z$ & Effective yield & $[0.0, 0.08]$\\
		$\tau$ & Gas infall timescale & $[0.0, 14.0]$Gyr\\
		$t_{0}$ & Start time of gas infall & $[0.0, 14.0]$Gyr\\
		$\lambda$ & The wind parameter & $[0.0, 10.0]$\\
		$t_{\rm cut}$ & The time that outflow turns off & $[0.0, 14.0]$Gyr\\
		$E(B-V)$&  Dust attenuation parameter & $[0.0, 0.5]$\\
		\hline
	\end{tabular}
\end{table}

\subsection{Spectral fitting approach}
\label{sec:fitting}
We explore the parameter space of the model 
described above to find a solution that matches the observations best. Such a fitting process is done through a Bayesian approach. In short, we first generate a set of model parameters from a prior distribution listed in \autoref{tab:paras}. Parameters related to the chemical evolution model are then used to predict corresponding star formation history (SFH) and chemical evolution
history (ChEH) following \autoref{eq:massevo}, \autoref{eq:cheevo1_MANGA} and \autoref{eq:cheevo2_MANGA}. We calculate a model spectrum from the predicted SFH and chemical following the standard stellar population synthesis procedure (see \citealt{Conroy2013} for a review). During the calculation, we use single stellar population (SSP) models from \cite{BC03} constructed using the STILIB empirical stellar spectra templates \citep{Borgne2003} with a Chabrier IMF \citep{Chabrier2003}. The SSP model spectra cover the range 3200--9500$\,${\AA} with a high spectral resolution of 3{\AA} and are available for metallicity values $Z$ from 0.0001 to 0.05 and ages from 0.0001$\,$Gyr to 20$\,$Gyr,
which are suitable for fitting MaNGA galaxies. We broaden the SSP templates to account for the broadening due to stellar velocity dispersion and instrumental effects in the observed spectra, which is inferred from an initial pPXF \cite{Cappellari2017} fit to the spectra. During this initial fit, emission lines in the observed spectra are identified and masked out in the subsequent analysis. Regarding dust attenuation, we simply use a screen dust model characterised by a \cite{Calzetti2000} attenuation curve. Finally, the model spectrum, as well as the current SFR and gas-phase metallicity, are compared with the observation through the following
$\chi^2$-like likelihood function: 
\begin{equation}
\label{likelyhood}
\ln {L(\theta)}\propto-\sum_{i}^N\frac{\left(f_{\theta,i}-f_{\rm D,i}\right)^2}{2f_{\rm err,i}^2}-
\frac{(Z_{\rm g,\theta}-Z_{\rm g,D})^2}{2\sigma_{Z}^2
}-\frac{(\psi_{0,\theta}-\psi_{\rm 0,D})^2}{2\sigma_{\psi}^2},\,
\end{equation}
where $f_{\theta, i}$ and $f_{\rm D, i}$ are fluxes at the $i$-th wavelength point from the model prediction and observed data respectively, with $f_{\rm err,i}$ being the corresponding error. The summation is made through all the $N$ wavelength points. Similarly, $Z_{\rm g,\theta}$ and $Z_{\rm g,D}$ are current gas-phase metallicities from the chemical evolution model and from the emission line diagnostics respectively, with  $\sigma_{Z}$ being the estimated uncertainty. The final term compared the predicted SFR from the model ($\psi_{0,\theta}$) to the observed value obtained from H$\alpha$ line emission ($\psi_{D,\theta}$), under an uncertainty $\sigma_{\psi}$. The Bayesian sampling was done with the {\tt MULTINEST} sampler \citep{Feroz2009,Feroz2019} and its \textsc{Python} interface \citep{Buchner2014}. When convergence is reached, we obtain the best-fit parameters from the posterior distribution, from which we calculate the SFH and ChEH of the galaxy and use them in the subsequent analysis.

\section{Result and discussion }
\label{sec:results}
Having obtained the best-fit galaxy properties for the galaxies, we are now in a position to investigate how the halo mass and stellar mass affect their evolution, including the galaxies' star formation and chemical enrichment histories.

\subsection{Global properties}
We start by exploring the global properties of each galaxy, obtained by fitting the observed data stacked within $1R_{\rm e}$. To separate the effects from halo mass and stellar mass, we divide the sample into small ranges of one property and see whether the other property has a significant impact on the evolution of these galaxies. 

\begin{figure*}
    \centering
    \includegraphics[width=1.0\textwidth]{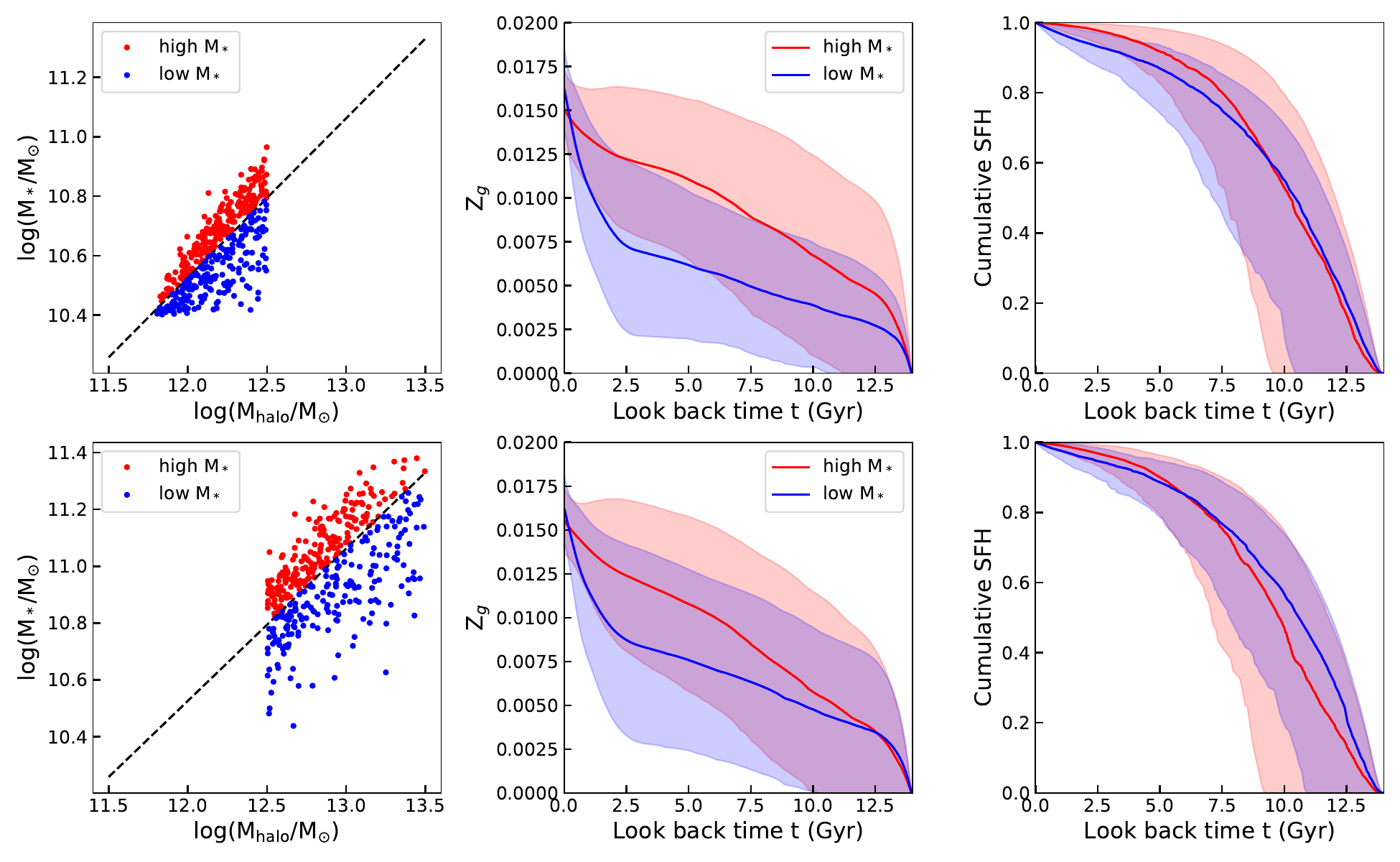}\\
     \caption{The effect of stellar mass on the evolution of massive central galaxies with halo mass in the range $10^{11.5} {\rm M}_{\odot} \leq M_{\rm h}\leq 10^{12.5} {\rm M}_{\odot}$ (top) and $10^{12.5} {\rm M}_{\odot} \leq M_{\rm h}\leq 10^{13.5} {\rm M}_{\odot}$ (bottom). In each row, the left panel shows high $M_{*}$ galaxies in red and low $M_{*}$ galaxies in blue, separated by the mean stellar mass-halo mass relation shown as a dashed line. In the middle (right) panel, solid lines in corresponding colours show the ChEH (cumulative SFH) averaged over the galaxies in each category, with shaded regions indicating the $1\sigma$ scatter.
     }
     \label{fig:fixhalo}
\end{figure*}

\subsubsection{The effect of stellar mass at fixed halo mass}

We first focus on the effect of the stellar mass of the galaxies, which has been known to be the main driver of the variation in the the galaxies' properties and evolution. As revealed by many previous investigations, galaxies generally follow a so-called `down-sizing' formation scenario -- more massive galaxies tend to accumulate their stellar masses and become chemical-enriched earlier than galaxies of lower stellar masses \citep[e.g.][]{Kauffmann2003,Gallazzi2005,Thomas2005}.  It is thus interesting to know whether such an effect holds when the halo mass of the galaxy is controlled for. To achieve this, we perform a linear fit to the stellar mass--halo mass relation (shown as dash lines in the left panels of \autoref{fig:fixhalo}). 
Note that in this plot we only include the 
sample of central galaxies, so that the halo mass can be regarded as the mass of the galaxies' own halo. 

At a given halo mass, galaxies that fall above the relation (denoted as high $M_{*}$ galaxies) are relatively more massive (in stars) than those that fall below the relation (denoted as low $M_{*}$ galaxies). We then obtain the averaged chemical enrichment history and stellar mass accumulation history of galaxies in the two categories and compare them with each other in \autoref{fig:fixhalo}. In the upper panels of \autoref{fig:fixhalo}, we focus on galaxies with halo masses in the range of $10^{11.5} {\rm M}_{\odot} \leq M_{\rm h}\leq 10^{12.5} {\rm M}_{\odot}$. It is clear that high $M_{*}$ galaxies (red) become chemically enriched systematically earlier than low $M_{*}$ galaxies (blue), even when they reside in halos of similar masses. Especially, high $M_{*}$ galaxies tend to build up their metal content very quickly in the first several Gyrs and then evolve slowly with time. On the other hand, the chemical enrichment processes of low $M_{*}$ galaxies tend to be suppressed at the beginning but then experience a secondary increase in the last few Gyrs. From  the results of paper I and those of an earlier investigation of galaxies at $z\sim0.7$ selected from the LEGA-C survey \citep{Zhou2024}, this kind of evolution may be be the consequence of AGN activity in the galaxies. Correspondingly, we see that high $M_{*}$ galaxies also tend to finish their star formation process earlier -- a larger fraction of stars in low $M_{*}$ galaxies form in the last $5\,$Gyrs, which leads to the recent increase in their metal content. 

Similar results can be found in the lower panels of \autoref{fig:fixstellar}, where we focus on galaxies living in more massive halos ($10^{12.5}  {\rm M}_{\odot} \leq M_{\rm h}\leq 10^{13.5} {\rm M}_{\odot}$), but the differences between the low- and high-$M_{*}$ categories appear smaller.

This result is very much consistent with, and reinforces, the long-established `down-sizing' paradigm. What is new here is that our results show that `down-sizing' still exists even if we control for halo masses; in other words, we conclude that `down-sizing' is not solely driven by halo mass evolution, but some internal baryonic process -- linked with stellar mass -- also play a role. 

Similar conclusions were reached by \cite{Scholz-Diaz2022} and \cite{Oyarzun2022} using average age and metallicities. However, our work provides a more detailed picture by obtaining the full history of the galaxies, showing that the different evolution between massive galaxies in the high and low $M_{*}$ categories can be traced back to more than $12\,$Gyrs ago. 

Note that \cite{Oyarzun2022} investigated mainly passive galaxies, whose star formation activity has been fully quenched for several Gyrs. In their results, the difference between the stellar population properties in high and low $M_{*}$ galaxies would have been in-printed by the star formation taking place at very early epochs in their evolution. Our work, centred on disk galaxies, complements that of \cite{Oyarzun2022}. Both pieces of research find similar differences in the early evolution of high- and low-$M_{*}$ galaxies which apply equally to present-day passive and disk galaxies. This suggests that the physics driving the evolution of these galaxies has been at play since very early times. We will come back to this point and discuss it in more detail in \autoref{sec:discus_stellarmass}.

\subsubsection{The effect of halo mass at fixed stellar mass}

\begin{figure*}
    \centering
    \includegraphics[width=1.0\textwidth]{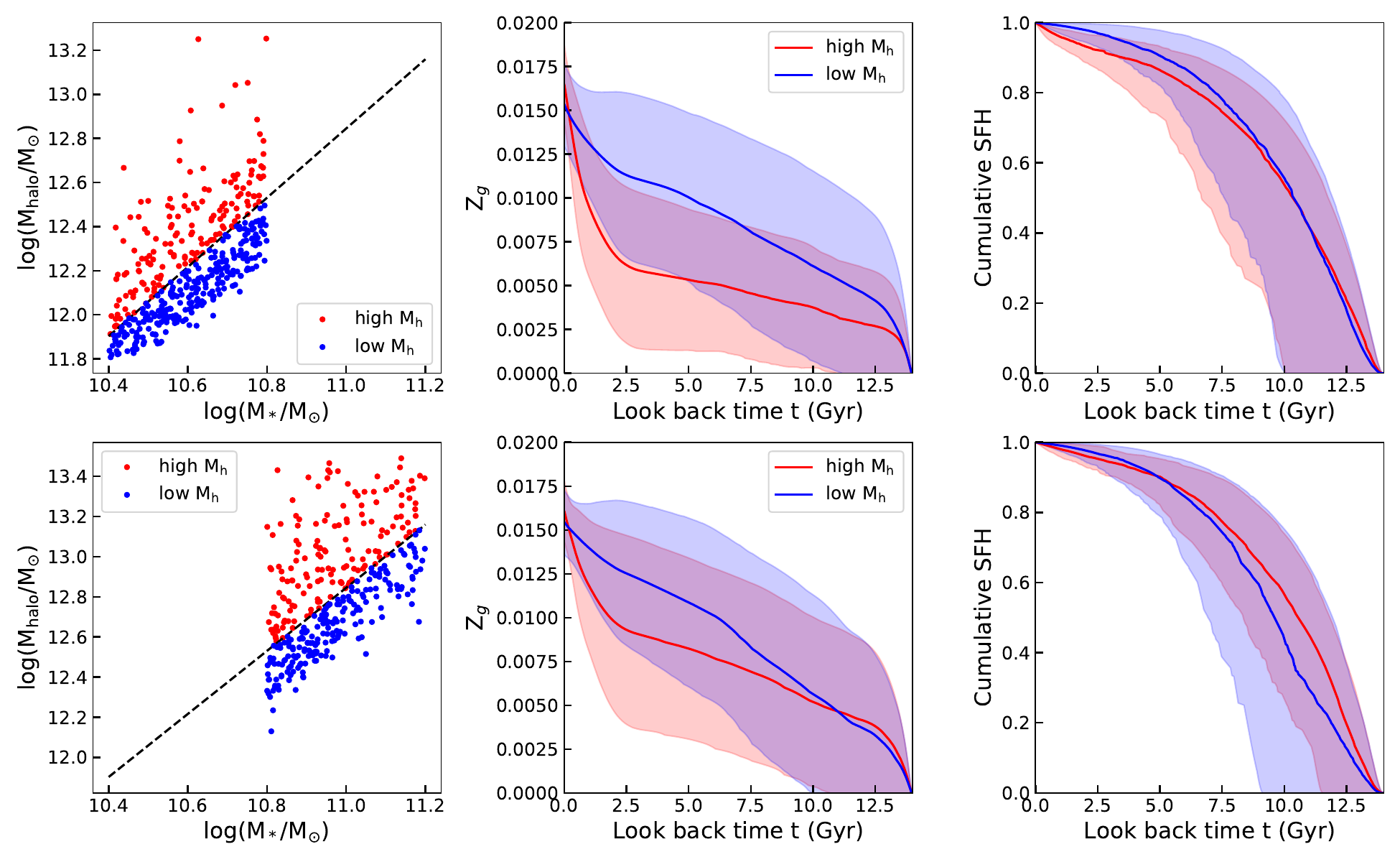}\\
     \caption{The effect of halo mass on the evolution of massive central galaxies at a given stellar mass. Plots are similar to \autoref{fig:fixhalo}, but the galaxies are divided into high $M_{\rm h}$ galaxies in red and low $M_{\rm h}$ galaxies using the mean halo mass-stellar mass relation.
     }
     \label{fig:fixstellar}
\end{figure*}

We now move on to see whether the halo mass has an impact on the galaxies' evolution when the stellar mass is controlled for. In parallel to what we did before, we now perform a linear fit to the halo mass--stellar mass relation, and use the relation to divide our galaxy sample into high and low halo-mass ($M_{\rm h}$) galaxies by selecting those who fall above and below the mean relation respectively. 

In the upper panels of \autoref{fig:fixstellar}, we
compare galaxies in the stellar mass range $10^{10.4} {\rm M}_{\odot} \leq M_{\rm *}\leq 10^{10.8} {\rm M}_{\odot}$ but with different halo masses. It is intriguing to see that galaxies seem to know how massive their halo is -- at fixed stellar mass a galaxy residing in a more massive halo tends to become chemically enriched later than those residing in lower-mass halos. High $M_{\rm h}$ galaxies also tend to have more recent star formation, as indicated by their later mass accumulation (right panel). 

Again, the results obtained from our complete star-formation and chemical enrichment histories echo the findings from averaged ages and metallicities presented in \cite{Scholz-Diaz2022} and \cite{Oyarzun2022}. 

Similar results hold for more massive galaxies ($10^{10.8} {\rm M}_{\odot} \leq M_{\rm *}\leq 10^{11.2} {\rm M}_{\odot}$) as shown in the lower panels of \autoref{fig:fixstellar}. In these more massive galaxies, the differences in metal enrichment histories between the high- and low-$M_{\rm h}$ galaxies also become smaller. In addition, there is a hint in the SFH that high $M_{\rm h}$ galaxies may have formed their stars earlier than low $M_{\rm h}$ ones. Such effect can also be seen for massive passive galaxies in  \cite{Oyarzun2022}, who found that the most massive ($M_{\rm *}> 10^{11} {\rm M}_{\odot}$) passive central galaxies in more massive halos can be older (albeit more metal-poor) than those in less massive halos.

\begin{figure*}
    \centering
    \includegraphics[width=1.0\textwidth]{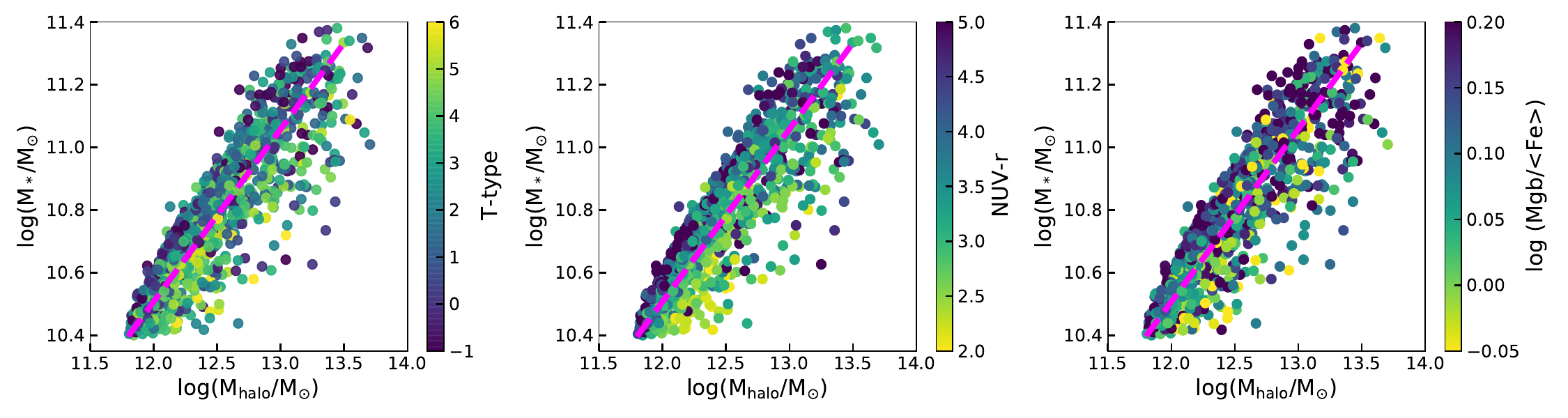}\\
     \caption{Distributions of morphologcal type (left), NUV-r colour (middle) and Mgb/$\langle$Fe$\rangle$ index ratio (right) of central galaxies in our sample on the stellar mass--halo mass plane. In each plot, the average stellar mass--halo mass relation is shown as a magenta dash-line as reference.}
     \label{fig:relation_1re}
\end{figure*}

\subsubsection{Empirical evidence}

The intriguing influence of stellar mass and halo mass on the evolution of galaxies described above is derived from our best-fit model results. To avoid possible model-dependent biases, we explore now alternative empirical evidence as a model-independent check.

In \autoref{fig:relation_1re} we investigate the distribution of morphological type (left), $NUV-r$ colour (middle), and Mgb/$\langle$Fe$\rangle$ index ratio (right) on the stellar mass--halo mass plane. On the left panel, the morphology of a galaxy is indicated by its T-type, with T-type$\,\leq0$ for early-type galaxies, T-type$\,>0$ for late-type galaxies, and T-type$\,=0$ for S0s. These T-types are obtained from one of the SDSS value-added catalogues (VAC), the MaNGA Morphology Deep Learning DR17 Catalogue \citep{Sanchez2022}, in which a Deep Learning approach is applied to all MaNGA galaxies to determine their morphology. We see a trend that seems to go perpendicular to the stellar mass--halo mass relation: galaxies that fall above the relation tend to have earlier morphologies. 

It is well known that galaxies with early morphological types tend to form their stars and quench their starf formation earlier than late-type galaxies, so it is not surprising to see the different evolutionary paths of galaxies in \autoref{fig:fixhalo} and \autoref{fig:fixstellar}. Such a trend is also seen clearly in the middle panel, in which the distribution of $NUV-r$ colour, a good indicator of recent star formation activity, is shown. Star-forming galaxies tend to have $NUV-r<4$, while quenched galaxies often have $NUV-r>4$. Galaxies that fall above the stellar mass--halo mass relation have lower recent star formation activity, which matches the expectations from their morphologies, as shown on the left panel.

The right panel of \autoref{fig:relation_1re} shows the Mgb/$\langle$Fe$\rangle$ index ratio, an indicator of the star-formation timescale. The Mgb index is a good tracer of the abundance of $\alpha$-elements, which are mostly released to the ISM by core-collapse supernovae. The Fe5270 and Fe5335 indices correlate with the abundance of iron, which is produced in both type Ia and core-collapse supernovae \citep{Nomoto1984}.
Since core-collapse supernovae explode very soon after star formation, while the progenitors of type Ia supernovae are relatively low-mass stars and therefore need a longer time to evolve, stellar populations formed in a short time-scale tend to have enhanced $\alpha$-element abundances,  indicated by higher  Mgb/$\langle$Fe$\rangle$ values. This is because their ISM had not been polluted by type Ia supernova explosions by the time  they stopped forming stars \citep{Thomas2005}. Galaxies located above the stellar mass--halo mass relation tend to have shorter star-formation timescales, as indicated by their higher  Mgb/$\langle$Fe$\rangle$ values. This empirical observation is in excellent qualitative agreement with the results obtained from the best-fit SFHs shown in the right panels of \autoref{fig:fixhalo} and \autoref{fig:fixstellar}.  

Alternative explanations have been proposed for the variation of the Mgb/$\langle$Fe$\rangle$ index ratio not directly related to the star-formation timescale. One of them is that galaxies with a shallower IMF slope at the high-mass end would show an enhancement in Mg, as there would be more massive stars that can release Mg to the ISM. In this scenario, our findings indicate that more massive galaxies hold a shallower IMF slope at the high-mass end. In fact, such a slope variation is supported by \cite{Worthey2011}, who find from the variation of [Ca/Fe] that more massive elliptical galaxies may have an IMF that favors more massive stars. However, it is somehow in conflict with some recent investigations of the IMF in galaxies -- at the low-mass end, studies based on either kinematics \citep[e.g.][]{Cappellari2012,Li2017} or small absorption features \citep[e.g.][]{vanDokkum2012,Navarroa2015,Zhou2019} in the spectra favor a bottom-heavy IMF in more massive galaxies. At the high-mass end, analysis on Wolf-Rayet stars also indicates a steeper slope in more massive or metal-rich galaxies \citep{Liang2021}.
A possible scenario that accommodates all the available observational evidence has recently been proposed by \cite{denBrok2024}, who speculate that the high-mass end IMF slope is representative of the very early age of the galaxies, and the low-mass end slope of the later star formation. In our current work, however, we are not able to explore the variation of IMF in our model due to the limited data quality. In what follows we will stick to the widely-accepted interpretation that the enhancement in Mgb/$\langle$Fe$\rangle$ or $\alpha$ element abundance indicates shorter ster-formation timescales.

\begin{figure}
    \centering
    \includegraphics[width=0.5\textwidth]{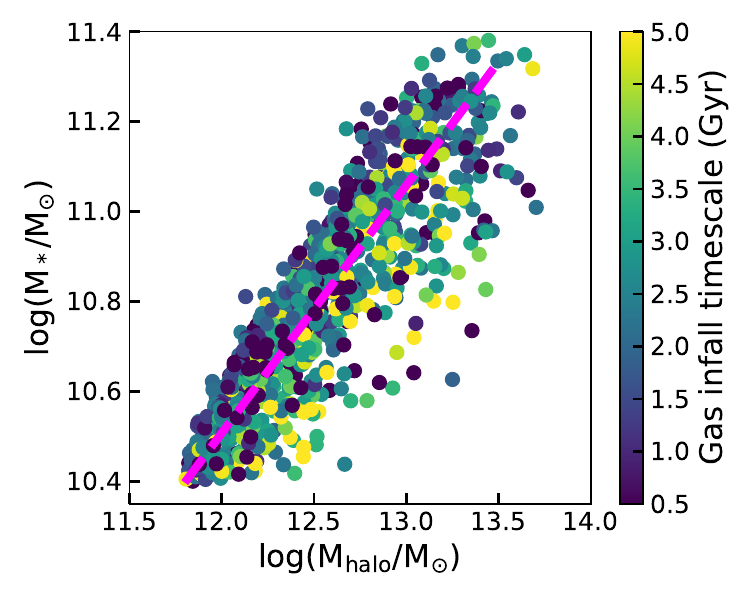}\\
     \caption{Distributions of gas infall timescale on the stellar mass--halo mass plane. The average stellar mass--halo mass relation is shown as a magenta line as reference.
     }
     \label{fig:gasinfall}
\end{figure}

In our fitting process, we use a stellar population model with fixed solar abundance patterns. In this case, the variation of the Mg and Fe indices themselves due to the changes of $\alpha$ abundance in the galaxies are not expected to affect the fitting results. However, we are able to investigate the effect of such variations in an indirect way -- in our model, the infall timescale of the gas is the main driver of the  star formation timescale of a galaxy, which is empirically linked to the variations in Mgb/$\langle$Fe$\rangle$ ratios. From the observed  Mgb/$\langle$Fe$\rangle$ ratios shown on the right panel of \autoref{fig:relation_1re}, we expect that, if our model fits are catching some real physics, the best-fit gas infall timescale should show a consistent distribution on the stellar mass--halo mass plane. To check this we use \autoref{fig:gasinfall}, which is similar to  the right panel of \autoref{fig:relation_1re} but now colour-coding galaxies by the best-fit gas infall timescale instead of Mgb/$\langle$Fe$\rangle$. Reassuringly, the results are just as expected: galaxies that lie above the stellar mass--halo mass relation have systematically shorter gas infall timescales.

All the consistent results we have found following different approaches support the scenario that both the total mass assembled \textit{and} the timescale of the mass assembly process have a significant impact on how a galaxy formed and evolved. We will come back to this point and discuss it in more detail in \autoref{sec:origins}.

\subsubsection{The influence of halo mass on the evolution of satellite galaxies}

\begin{figure*}
    \centering
    \includegraphics[width=1.0\textwidth]{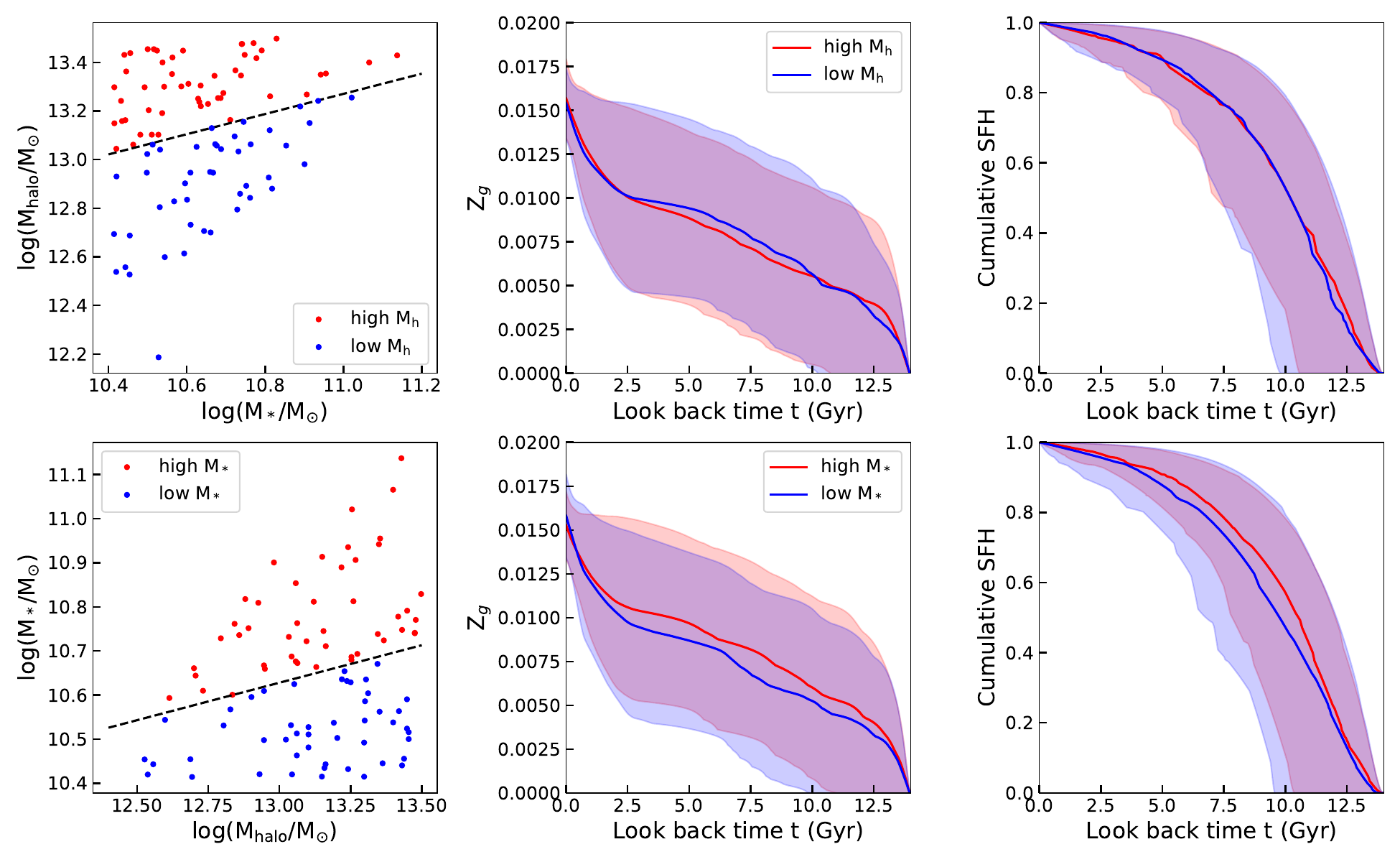}\\
     \caption{The effect of group halo mass (top) and stellar mass (bottom) on the evolution of massive satellite galaxies. The left panels show how we separate our sample into two sub-samples. In the middle (right) panel, solid lines in corresponding colours show the ChEH (cumulative SFH) averaged over galaxies in each subsample, with shaded regions indicating the 1$\sigma$ scatter.
     }
     \label{fig:satellite}
\end{figure*}

As a comparison, we also check whether the halo mass has a significant impact on the evolution of massive satellite galaxies. Note that, for satellite galaxies, the halo mass here is the mass of the group/cluster halo they inhabit instead of the individual halo of the galaxy itself. In the top panels of \autoref{fig:satellite}, we divide the satellite sample into high $M_{\rm h}$ and low $M_{\rm h}$ galaxies following a similar approach to that used in \autoref{fig:fixstellar}. Contrary to what we found for central galaxies, we do not observe any difference between the high- and low-$M_{\rm h}$ samples, indicating that the group halo mass has little impact on the evolution of these massive satellite galaxies. 
Conversely, in the bottom panels of \autoref{fig:satellite} we control for the group halo mass and divide our sample into high $M_{\rm *}$ and low $M_{\rm *}$ galaxies, from which we see that the stellar mass still plays some role in shaping the evolution of the galaxies, although the impact is small due to the coarseness of the division and the relatively small sample size. 

This result is not surprising since it is well known that the influence of the group/cluster environment on satellites is mostly limited to low-mass galaxies \citep[e.g.][]{Pasquali2010,Wetzel2012,Peng2012,Zhou2022environment}. For the relatively high mass galaxies in our sample, their potential wells are often deep enough to protect the gas in the galaxies from being stripped out by the environment, and the environmental effect is therefore weaker. Given the results, in what follows we will only focus on the central disk galaxies in our sample  and try to figure out the possible physical drivers of the observed trends.

\begin{figure*}
    \centering
    \includegraphics[width=1.0\textwidth]{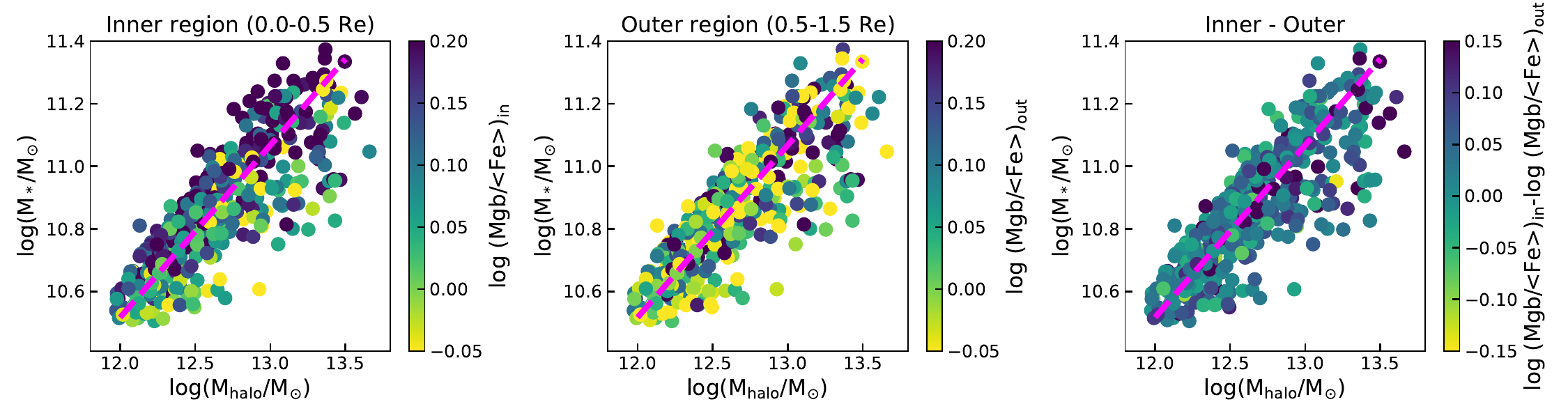}\\
    \includegraphics[width=1.0\textwidth]{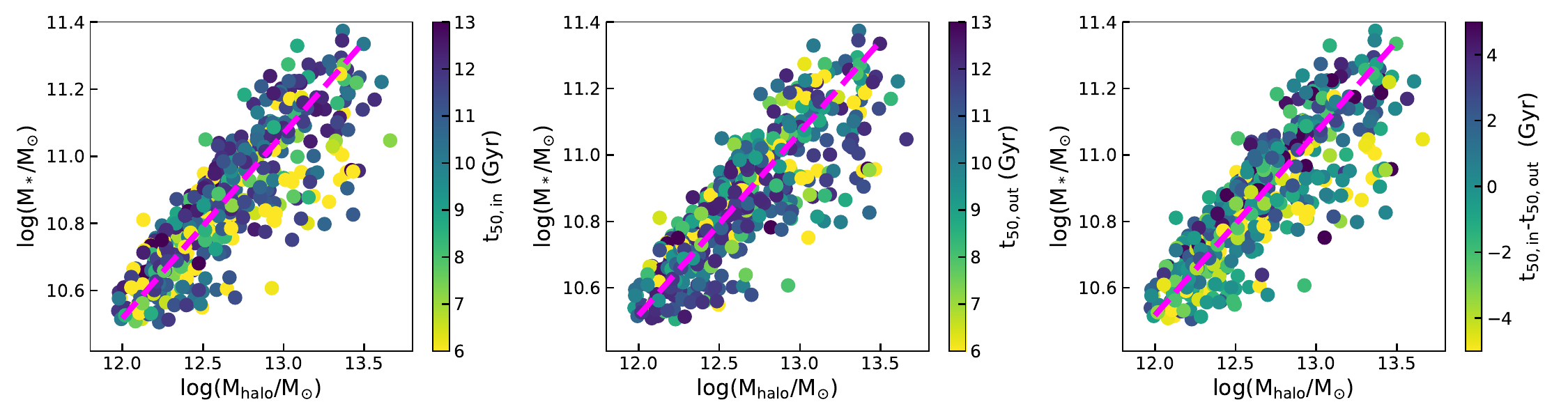}\\
    \includegraphics[width=1.0\textwidth]{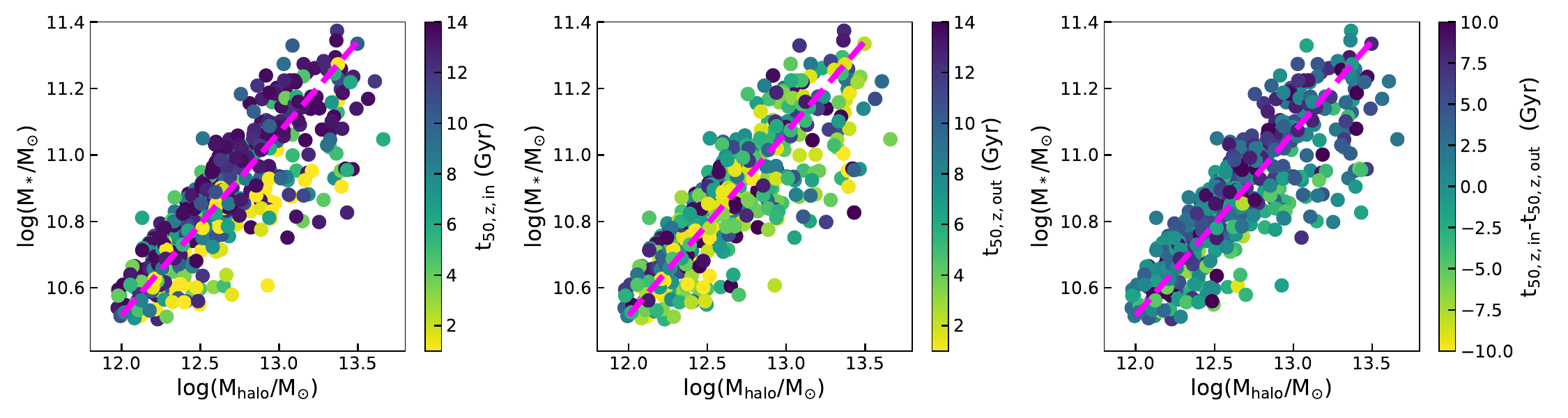}\\
     \caption{Distributions of the Mgb/$\langle$Fe$\rangle$ index ratio (first row),  half mass formation time ($t_{50}$, second row) and half metallicity enrichment time ($t_{50,Z}$, third row) on the stellar mass--halo mass plane. In each row, the left panel shows properties obtained from the inner region (0.0--0.5$\,R_{\rm e}$) of the sample galaxies, the panel in the middle shows results from the outer regions (0.5--1.5$\,R_{\rm e}$), with the right panel showing the differences between the inner and outer regions. In each plot, the average stellar mass--halo mass relation is shown as a magenta line as reference.
     }
     \label{fig:relation_nre}
\end{figure*}

\subsection{Additional evidence from spatially-resolved results}

From the global properties analysied above, we have learned that both halo mass and stellar mass have an impact on the evolution of central disk galaxies, while for satellite disk galaxies only the stellar mass of the galaxy plays a role. From our best-fit models, the impact is very likely to act by changing the gas infall timescale of the galaxy. If this scenario is correct, we may naturally expect that such a mechanism leaves footprints mainly at the centre of the galaxy. The reason for such expectation is simple: current galaxy formation models suggest that these massive halos mainly grow through major mergers that happened mostly before $z\sim2$\, and the majority of the stars in the centre of the galaxies often formed during the violent gas infall and star formation triggered by such events \citep[e.g.][]{Mo1998}. In contrast, the outer parts of the galaxy, especially the outer disk, often formed through long-timescale secular evolution processes or minor mergers, which would not be expected to retain much memory of the halo formation or the gas infall histories of the galaxy a long time ago.

The spatially-resolved information provided by MaNGA gives us a great opportunity to test such a scenario. In this section, we focus on how the evolution of central galaxies can be affected by the halo mass and stellar mass in a spatially resolved way. Specifically, we divide each galaxy into inner (0.0--0.5$\,R_{\rm e}$) and outer (0.5--1.5$\,R_{\rm e}$) regions, and investigate the distributions of the resolved galaxy properties on the stellar mass--halo mass plane. 

In the first row of \autoref{fig:relation_nre}, we show the distribution of Mgb/$\langle$Fe$\rangle$ on the stellar mass-halo mass plane for the central (left panel) and outer (panel in the middle) regions respectively, and in the right panel we show the difference between values in the two regions. We see that  Mgb/$\langle$Fe$\rangle$ shows a trend in the central region that is similar to that found for the global values (shown in the right panel of \autoref{fig:relation_1re}): galaxies that lie above the stellar mass--halo mass relation tend to have higher  Mgb/$\langle$Fe$\rangle$. As discussed above, this higher Mgb/$\langle$Fe$\rangle$ indicates shorter star formation timescales, which may linked to faster assembly of their gas content. In contrast, in the outer regions (middle panel) we do not observe a similar trend; the outer regions of more massive galaxies tend to have higher  Mgb/$\langle$Fe$\rangle$ values, but these show no correlation with the halo mass. This result suggests that  the outer region of a galaxy is not ``aware'' of the mass of the halo. The difference between the Mgb/$\langle$Fe$\rangle$ values in the inner and outer regions (i.e., the radial gradient) is shown on the right panel; no clear trend is found there either. Such a result is consistent with the proposed scenario, confirming that it is the evolution of the central region of the galaxy which is affected by both the stellar mass and the halo mass of the galaxy.

We further test such a scenario using our semi-analytic spectral fitting approach. We apply our models to data stacked within the inner and outer regions of the galaxies (see \autoref{sec:data}). From the best-fit parameters, we obtain the best-fit star formation and chemical evolution histories for the two regions in each galaxy.  In the second row of \autoref{fig:relation_nre}, we make plots similar to those in the first row, but now for the half-mass formation time $t_{50}$ obtained from the best-fit SFHs of the galaxies. $t_{50}$ indicates the time when the region accumulates half of its stellar mass, which also reflects the timescales of the star formation in the region. Similar to the top panels, we see an even clearer trend in the central region of the galaxy (left), and the absence of the correlation for the outer region  (middle) and the gradients (right).  
Similarly, in the third row of \autoref{fig:relation_nre} we present a plot for $t_{50, Z}$, which is defined as the time when the region's metallicity reaches half of its present-day value.
We see very similar trends in $t_{50, Z}$: in the central region of the galaxies $t_{50, Z}$ shows a systematic trend perpendicular to the stellar mass--halo mass relation, while in the outer regions, the trend becomes much weaker. Such a result indicates again that the gas infall timescale or halo formation process leave footprints mainly in the central region of the galaxies.

We end the discussion by comparing our spatially-resolved results with those of previous works.  \cite{Scholz-Diaz2022} use single-fibre SDSS spectra to investigate the scatter of galaxy properties across the stellar-to-halo mass relation. Although they only consider averaged ages and metallicities, their results are very similar to the ones we find from the spectra stacked within $1R_{\rm e}$ or the central $0.5R_{\rm e}$. Considering the 3 arcsecond SDSS fibre diameter, the galaxy properties they obtained are mostly from the central regions, so it is not surprising to find consistent results for the centre of the galaxies. Such a good agreement between different approaches and datasets gives us further confidence that we are capturing some real physics. 

\cite{Oyarzun2022} use similar IFU data from MaNGA to explore how stellar populations depend on stellar and halo mass for passive central galaxies. They report that at fixed stellar mass, the evolution of these galaxies has a significant dependence on the halo mass, and the dependence holds until $1.5R_{\rm e}$. 
At first glance, their spatially resolved results seem to be inconsistent with what we find for the outskirts of the galaxies. However, their sample contains only passive galaxies, which means that their star formation activity was already quenched several Gyrs ago. Without being polluted by the stellar populations formed recently, which do not have memory of the halo formation in the distant past, it can be expected that the record of the halo formation can be found in these passive galaxies at all radii. In fact, both results support the idea that it is the stellar populations formed in the early universe that ``know'' the properties of the dark matter halo. 

In what follows, we will discuss in detail the possible physical origin of the trends we have observed.

\subsection{Possible physical origin of the observed trends}
\label{sec:origins}

\subsubsection{The effect of halo mass on the evolution of central galaxies}

In the current galaxy formation paradigm, galaxies form at the centre of dark matter halos, so their evolution ought to be deeply affected by the growth and evolution of the dark matter halos themselves. It has thus long been proposed that, as the main parameter that determines the evolution of a dark matter halo, its mass may also drive the growth and evolution of the galaxy at its centre \citep[e.g.][]{White1978,Blumenthal1984,Mo1998}. 
Indeed, we do observe a correlation between halo mass and the stellar population
properties of central galaxies even when the stellar mass of the galaxy is controlled for; when considering galaxies of similar stellar mass, those with higher halo masses tend to be late in their chemical enrichment and thus have lower averaged stellar metallicity, and they also have more recent star formation activity.

There are several possible physical processes that may be causing such trends. A natural explanation is that,  
at fixed stellar mass, more massive halos are more efficient at feeding gas into their central galaxy. The infalling pristine gas can dilute the ISM in the central galaxy, making it more metal-poor than that of galaxies in less massive halos. In addition, the long-lasting gas infall provides material for more recent star formation. Such a scenario is supported by simulations. \cite{Wang2023} found from the EAGLE simulations that, at a given stellar mass, the ISM fraction of central galaxies in more massive dark matter halos at high redshift is generally higher than that in less massive ones. Our best-fit model results suggest longer gas-infall timescales in high $M_{\rm h}$ galaxies, which is also evidence of more efficient gas accretion.

Alternatively, it is thought that in these massive galaxies AGN feedback can play an important role \citep[e.g.][]{Croton2006,Sijacki2007} in shaping their evolution. Strong AGN feedback can remove metal-enriched gas from the galaxy and suppress its chemical enrichment. Therefore, it is also possible that the halo properties correlate with the activity of the central black hole at a given stellar mass so that this connection can shape the evolution of the galaxy through feedback mechanisms. From the EAGLE simulations, \cite{Wang2023} did find that, 
at a given stellar mass, central galaxies in more massive dark matter halos tend to have high black hole masses, which they speculate it may be another mechanism for halos to affect the stellar population properties of galaxies. However, using a variety of observational evidence, \cite{Kormendy2013} conclude that the halo mass of a galaxy is not likely to be more fundamentally correlated with the central black hole mass than stellar mass. In this case, it is still in doubt whether the halo shapes the evolution of the galaxy by altering the properties and activity of the central supermassive black hole.

\subsubsection{The effect of stellar mass on the evolution of central galaxies}
\label{sec:discus_stellarmass}
Our results show that the stellar mass of the galaxy is crucial in regulating the stellar population properties even at fixed halo mass. At a given halo mass, galaxies with higher stellar masses tend to be more metal-rich and host older stellar populations. However, in the current galaxy formation framework, a galaxy’s growth history ought to be ultimately tied to the formation and evolution of its
dark matter halo. In that case, other halo
parameters beyond its mass $M_{\rm h}$ (especially those which correlate with the stellar mass of the galaxy), should have played an important role. However, it is often hard to infer the properties of dark matter halos beyond their mass, nor to correlate them with stellar population properties such as age and metallicity, as dark matter halos can not be observed directly.

In these circumstances, people often seek solutions using semi-analytic models or hydrodynamic simulations of galaxy formation and evolution, trying to find other halo properties that may possibly be linked to the galaxies' stellar populations. Promising progress along these lines has been made by \cite{Croton2007}, who found from semi-analytic models that the halo formation time correlates with galaxy luminosity at fixed
halo mass. Since then, various hydrodynamic simulations and other semi-analytical models have confirmed that, at fixed halo mass, earlier-formed halos host more massive galaxies than later-formed ones \citep[e.g.][]{Wang2013,Zehavi2018,Martizzi2020}. We can thus speculate that our findings of a dependence of the galaxy properties on stellar mass at fixed halo mass may reflect the various formation times of the halos themselves.

In fact, such an idea is supported by the research of \cite{Scholz-Diaz2022} and \cite{Oyarzun2022}, who searched for variations in the galaxy properties with stellar mass while controlling for halo mass. In our own results we do find that the Mgb/$\langle Fe\rangle$ index ratio is generally higher for galaxies that fall above the stellar mass--halo mass relation, which indicates a faster accumulation of stellar mass in these galaxies. In combination with the shorter gas infall timescales obtained through our semi-analytic spectral fitting approach (see \autoref{fig:gasinfall}), we can naturally speculate that these galaxies reside within dark matter halos that formed systematically earlier than their low stellar mass counterparts. In addition, the 
 spatially resolved results shown in \autoref{fig:relation_nre} provide further evidence for such a scenario: galaxies that have higher stellar-mass to halo-mass ratio formed half of their stellar masses earlier as indicated by $t_{50}$, but this effect is only prominent in the central region of the galaxies. In the outer parts, there is little evidence that $t_{50}$ correlates with either stellar mass or halo mass. As stars in the central regions of these massive galaxies are generally older than those in the outer parts, this result indicates that it is those old stars the ones keeping memory of their dark matter halo formation process. The late-formed stars in the outer parts don't know much about the time the halo formed since the halo formation had almost finished several Gyrs before their formation. Therefore, the properties of the stars in the outer regions show no correlation with either halo mass or halo formation time.  It is worth pointing out that, since both the gas infall and star formation timescales are not linked to the halo formation timescales in a straightforward way, future research that including detailed hydrodynamic simulations is needed to provide more direct physical insights.

\section{Summary}
\label{sec:summary}
In this work we analyse a sample of massive disk galaxies selected from the SDSS-IV/MaNGA survey to investigate the dependence of the galaxies' evolution on their stellar and halo masses. We selected central galaxies as the most massive members of their group. A sample of satellite galaxies with similar stellar masses was also selected for comparison. For each galaxy, we combine the spectral data within one effective radius $R_{\rm e}$ to create a high signal-to-noise-ratio spectrum representing its global properties. Data inside the 0.0--0.5$\,R_{\rm e}$ and 0.5--1.5$\,R_{\rm e}$ regions are also combined to characterise the stellar populations in the central and outer regions of the galaxy. We applied a semi-analytic spectral fitting approach to the combined data to obtain key physical properties of the galaxies. This approach produces the star-formation and chemical-evolution histories from the galaxies' data and also determines physical parameters related to gas accretion and outflows, which are key components of the models. From the best-fit model results, as well as from direct observables such as morphologies, $NUV-r$ colours, and Mgb/$\langle$ Fe$\rangle$ index ratios, we obtained the following results:

\begin{itemize}
    \item We confirm the canonical `down-sizing' formation scenario in which more massive galaxies tend to accumulate their stellar mass and become chemically enriched earlier than less massive ones. This effect holds even when the mass of the dark matter halo is controlled for. 

    \item  In addition to the stellar mass,  halo mass is also found to have a significant impact on the stellar 
    population properties of the galaxies. At fixed stellar mass, galaxies in low-mass halos are found to be more metal-enriched and finish star formation earlier than those in more massive halos.

    \item The results based on the semi-analytic fitting process are entirely consistent with those based on direct observables such as morphological type, $NUV-r$ colour, and Mgb/$\langle$ Fe$\rangle$ index ratio of the galaxies. In particular, the star formation timescales indicated by Mgb/$\langle$ Fe$\rangle$ are in line with the gas infall timescales obtained through the best-fit chemical evolution models. All these results confirm the correlations with stellar and halo mass described above. 
    
    \item The halo mass of the host group/cluster is found to have little impact on the evolution of massive satellite galaxies at a given stellar mass. Instead, the evolution of massive satellite galaxies is mainly driven by their stellar masses.

    \item Using spatially resolved data from the inner and outer regions of the galaxies, we find that only the properties of the central regions have a dependence on halo mass. The outer regions of the galaxies seem to know little about the properties of their halos.

\end{itemize}

Based on these results, we discuss the possible physical origins of the dependencies we find. We conclude that at a given stellar mass, more massive halos may be more efficient at feeding gas into the central galaxy so that their star formation is long-lasting. This gas inflow also dilutes the chemical composition of the ISM in these galaxies, slowing down their chemical enrichment. At a given halo mass, a galaxy with higher stellar mass may live in a halo formed earlier, so that their star formation activity also quenches earlier. These simple physical ideas, backed by more sophisticated hydrodynamic and semi-analytic galaxy evolution models, are able to explain the trends we observe.

\section*{Acknowledgements}
SZ, AAS and MRM acknowledge financial support from the UK Science and Technology Facilities Council (STFC; grant ref: ST/T000171/1).

For the purpose of open access, the authors have applied a creative commons attribution (CC BY) to any journal-accepted manuscript.

Funding for the Sloan Digital Sky Survey IV has been provided by the Alfred P. 
Sloan Foundation, the U.S. Department of Energy Office of Science, and the Participating Institutions. 
SDSS-IV acknowledges support and resources from the Center for High-Performance Computing at 
the University of Utah. The SDSS web site is www.sdss.org.

SDSS-IV is managed by the Astrophysical Research Consortium for the Participating Institutions of the SDSS Collaboration including the Brazilian Participation Group, the Carnegie Institution for Science, Carnegie Mellon University, the Chilean Participation Group, the French Participation Group, Harvard-Smithsonian Center for Astrophysics, Instituto de Astrof\'isica de Canarias, The Johns Hopkins University, Kavli Institute for the Physics and Mathematics of the Universe (IPMU) / University of Tokyo, Lawrence Berkeley National Laboratory, Leibniz Institut f\"ur Astrophysik Potsdam (AIP), Max-Planck-Institut f\"ur Astronomie (MPIA Heidelberg), Max-Planck-Institut f\"ur Astrophysik (MPA Garching), Max-Planck-Institut f\"ur Extraterrestrische Physik (MPE), National Astronomical Observatories of China, New Mexico State University, New York University, University of Notre Dame, Observat\'ario Nacional / MCTI, The Ohio State University, Pennsylvania State University, Shanghai Astronomical Observatory, United Kingdom Participation Group, Universidad Nacional Aut\'onoma de M\'exico, University of Arizona, University of Colorado Boulder, University of Oxford, University of Portsmouth, University of Utah, University of Virginia, University of Washington, University of Wisconsin, Vanderbilt University, and Yale University.

\section*{Data availability}
The data underlying this article were accessed from: SDSS DR17 \url{https://www.sdss.org/dr17/manga/}. The derived data generated in this research will be shared on request to the corresponding author.

\bibliographystyle{mnras}
\bibliography{szhou} 

\bsp	
\label{lastpage}
\end{document}